\documentclass[%
 reprint,
superscriptaddress,
 amsmath,amssymb,
 aps,
floatfix,
]{revtex4-2}

\usepackage{graphicx}
\usepackage{dcolumn}
\usepackage{bm}
\usepackage{natbib}
\usepackage{ulem}
\usepackage{xcolor}
\usepackage{amsmath} 
\usepackage{amssymb}
\usepackage{hyperref}
\usepackage{multirow}

\definecolor{AMK}{RGB}{0, 0, 200}

\begin{document}

\preprint{APS/123-QED}

\title{
Role of spatiotemporal nonuniformities in laser‑induced \\magnetization precession damping}

 \author{P. I. Gerevenkov}
\email{petr.gerevenkov@mail.ioffe.ru}
\homepage{http://www.ioffe.ru/ferrolab/}
\affiliation{Ioffe Institute, 194021 St. Petersburg, Russia}
\author{Ia. A. Filatov}
\affiliation{Ioffe Institute, 194021 St. Petersburg, Russia}
\author{L. A. Shelukhin}
\affiliation{Ioffe Institute, 194021 St. Petersburg, Russia}
\author{P. A. Dvortsova}
\affiliation{Ioffe Institute, 194021 St. Petersburg, Russia}
\author{A. M. Kalashnikova}
\affiliation{Ioffe Institute, 194021 St. Petersburg, Russia}

\date{\today}

\begin{abstract}
Laser-induced magnetization precession measurements in ferromagnets often reveal an anomalous decrease in the damping time near a field-induced second-order spin-orientation transition, a behavior that cannot be described by the linearized Landau-Lifshitz-Gilbert equation.
Here we demonstrate that this anomaly is not a material property but results from interference of precessing local magnetizations within the inhomogeneously excited region.
By combining pump-probe experiments, analytical modeling that accounts for the finite sizes of the pump and probe spots, and micromagnetic simulations, we show that the standard macrospin approach fails to capture the observed dynamics.
The inhomogeneous relaxation of magnetic parameters within the excitation area distorts the measured precession envelope, while dipole fields give rise to a temporally non-monotonic term in its frequency.
Our results highlight the critical role of excitation locality in a vicinity of critical fields.

\end{abstract}

\maketitle

The introduction of ultrafast laser excitations into magnetism research~\cite{Vaterlaus1991,Beaurepaire1996,vanKampen2002,Kimel2004} gave rise to the field now known as femtromagnetism.
Relaxation processes following femtosecond laser excitation have been of central interest since the earliest works~\cite{Beaurepaire1996,Kimel2004}, both for fundamental reasons and for potential applications in ultrafast information recording~\cite{Vahaplar2009}.
Since many femtomagnetic switching phenomena involve magnetization precession~\cite{Hanseen2025,dejong2012,afanasiev2016,stupakiewicz2017,schlauderer2019,davies2019,Kuzikova2025}, its damping is a key parameter that must be either enhanced or suppressed depending on the intended goal.
Several experiments in ferro- and ferrimagnets report an anomalous decrease in the laser‑induced precession damping time~\cite{walowski2008intrinsic,chen2012spin,Mondal2018}, particularly near spin‑orientation transitions~\cite{capua2015determination,davies2019,Frej2023}.
This anomaly has been observed in both metals and dielectrics and with both thermal~\cite{de2024coherent,solovev2021second} and non‑thermal~\cite{Frej2023} excitation of precession.
Various explanations have been proposed: anisotropy recovery on a timescale of approximately half the precession period~\cite{davies2019}; angular momentum compensation in ferrimagnets~\cite{stanciu2006ultrafast,xu2026inversion}; multimagnon scattering~\cite{capua2015determination,de2024coherent,rzhevsky2007magnetization}; and inhomogeneity of the precession frequency or excitation torque within the detection area~\cite{chen2012spin,walowski2008intrinsic,gareev2025strongly,mishra2023dynamics}.
At the same time, several authors note a discrepancy between the damping time deduced from laser‑induced experiments and the width of the microwave absorption line~\cite{capua2015determination,de2024coherent}.
Moreover, a qualitative disagreement exists between experimental data and solutions of the linearized Landau–Lifshitz–Gilbert (LLG) equation when describing laser‑induced precession damping.
Consequently, a reassessment of the evaluation of magnetization precession damping is required.

In this Letter, we report on time‑resolved magneto‑optical measurements of laser‑induced magnetization precession in an epitaxial Fe film with low intrinsic damping. 
We analyze the precession damping time and its frequency evolution tracked using windowed FFT analysis.
Analysis based on several models -- from a linearized LLG macrospin to full micromagnetic simulations -- reveals that the apparent reduction in damping time near the field‑induced spin‑orientation transition is fictitious to a great extent. 
It arises primarily from the interference of precessing local magnetizations across the area excited by a laser pulse with a Gaussian spatial profile.
Furthermore, we isolate the contribution of dipole fields and show that their temporal evolution is non‑monotonic, which leads to a significant underestimation of the laser‑induced heating when this contribution is neglected.

The experiments were carried out using an epitaxial Fe~(001) film with a thickness $d = 20$~nm grown by pulsed laser deposition on a MgO~(001) substrate~\cite{suturin2022laser,dvortsova2022technological}.
The studied film possesses cubic magnetocrystalline anisotropy with an easy axes along the $\langle$110$\rangle$ MgO directions and strain-induced anisotropy with the (001) easy-plane~\cite{khodadadi2020conductivitylike,tournerie2008plane,solano2022spin}.
Single-crystalline iron was chosen because it is a model magnonic material with a single magnetic sublattice, demonstrating laser-induced anisotropy relaxation times much longer than the precession period and long-lived magnetization dynamics~\cite{filatov2022spectrum,gerevenkov2021effect}.
Magnetization dynamics was tracked in time-resolved pump-probe measurements.
Femtosecond pump pulses were focused into a spot of 39~$\mu$m FWHM.
The probe pulses FWHM was approximately two times smaller (23~$\mu$m), which is usually assumed to be sufficient to consider uniform pump-induced heating within the probed area~\cite{chen2012spin,Frej2023}.
The external field $H_\mathrm{ext}$ was applied in the sample plane at an angle $\phi$ from one of the easy axes.
Micromagnetic simulations were performed using the GPU-accelerated micromagnetic framework mumax$^3$~\cite{vansteenkiste2014design}.

\begin{figure*}
\includegraphics[width= 0.9\linewidth]{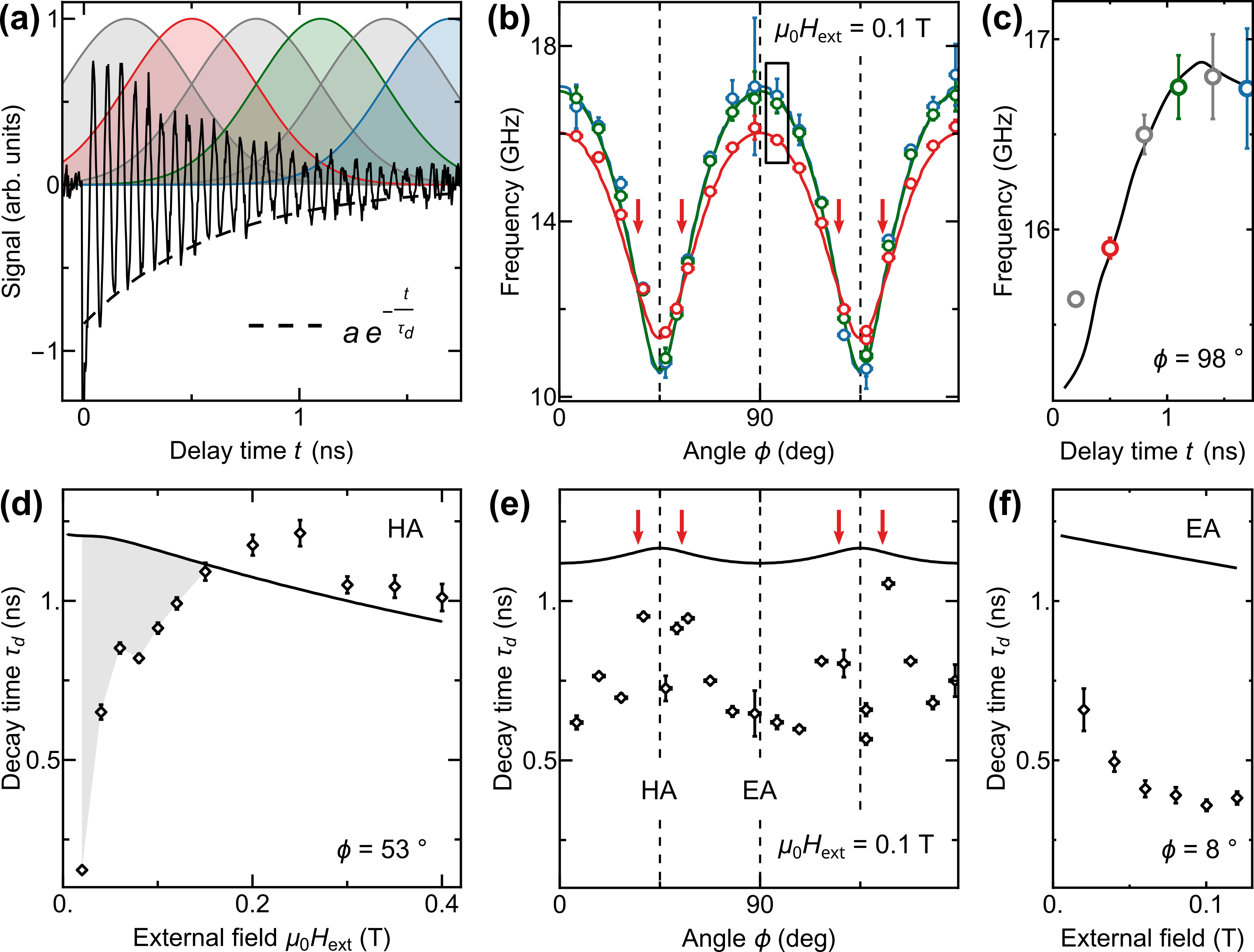}
\caption{\label{fig:uniform}
{\bf Azimuthal and field-strength dependences of laser-induced precession frequency and damping time.}
(a) Exemplary laser-induced precession of magnetization measured in an Fe film (black solid line).
Fourier windows used to obtain the frequency at different times are shown by colored Gaussians.
The dashed line is a fit to extract the initial amplitude and damping time.
(b) Points show the azimuthal dependence of the precession frequency at different times after excitation.
Lines show the corresponding fits using the Smit-Suhl approach.
Colors correspond to the time windows in panel (a).
(c) Central frequencies obtained with windowed FFT using the time windows from panel (a) and model prediction (line).
(d-f) Experimental damping time values and model predictions: (e) azimuthal angle $\phi$ dependence at $\mu_0 H_{ext} = 0.1$\,T; (d) and (f) external field dependencies at $\phi = 53$ and 8\,$^\circ$, respectively.
}
\end{figure*}

The standard approach to determining magnetic parameters in thin films relies on measuring the magnetization precession frequency as a function of the in-plane direction and strength of the external magnetic field. 
In pump-probe experiments, however, the magnetic parameters are not static; they evolve in time as the system relaxes from the heating caused by the laser-induced ultrafast excitation~\cite{gerevenkov2021effect}. 
Capturing this dynamics cannot be guided by a single frequency value.
In Fig.~\ref{fig:uniform}\,(a) we show a typical pump-probe signal obtained in experiment at excitation fluence $F = 28$\,mJ cm$^{-2}$, $H_\mathrm{ext} =  0.1$\,T, and $\phi= 98$\,$^\circ$, revealing a long-living laser-induced magnetization precession. 
We analyze such a time trace using a windowed fast Fourier transform (FFT).
We apply Gaussian windows with a standard deviation of $\sigma = 0.3$~ns, centered at 0.2 -- 1.7\,ns after the pump impact [color-coded Gaussians in Fig.~\ref{fig:uniform}\,(a)]. 
This allows us to track how the precession frequency evolves as the magnetic system cools down [Fig.~\ref{fig:uniform}\,(b,c)].

Using this approach, we map the frequency of laser-induced precession as a function of time [points in Fig.~\ref{fig:uniform}\,(c)], in-plane direction  [points in Fig.~\ref{fig:uniform}\,(b)], and strength (not shown) of the external magnetic field $H_\mathrm{ext}$. 
The azimuthal dependencies, measured at each temporal window, clearly reveal the cubic magnetic anisotropy of the Fe film: the frequency reaches maxima when the field is aligned with the easy axes (EA, along the $<$110$>$ directions of MgO) and minima when aligned with the hard axes (HA, along the $<$100$>$ directions of MgO). 
More importantly, the time-resolved measurement reveals a distinct evolution of the frequency as the system relaxes. 
The relaxation of the saturation magnetization $M_S$ and the cubic anisotropy constant $K_C$ toward their equilibrium values $M_S (\infty)$ and $K_C (\infty)$ drives a shift in the precession frequency. 
To keep the analysis simple, we omit the change of $K_U$.
The magnitude of this shift reaches nearly 1.5\,GHz, as illustrated for the near-EA direction of $\mathbf{H}_\mathrm{ext}$ by the points in Fig.~\ref{fig:uniform}\,(c). 
This temporal evolution underscores the need to go beyond static models and account for the dynamic magnetic response.

\begin{figure*}
	\includegraphics[width= 0.9\linewidth]{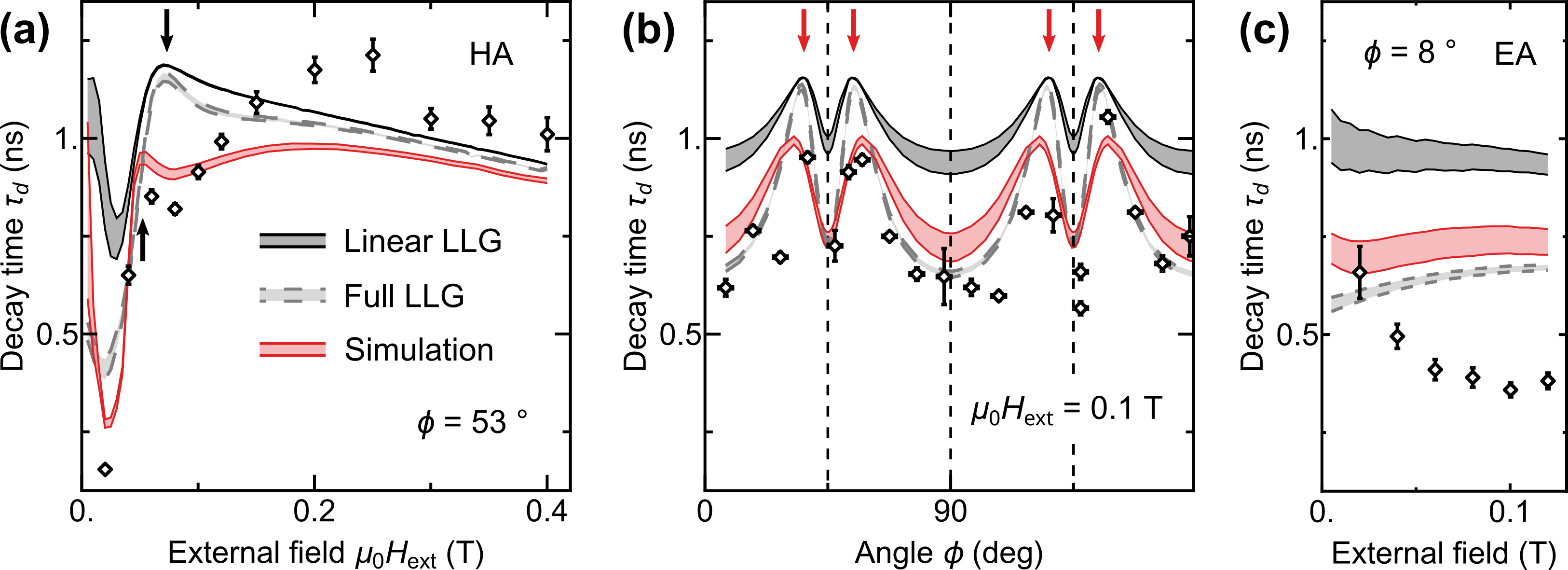}
	\caption{\label{fig:nonuniform}
		{\bf Analysis using laterally nonuniform linearized and full LLG.}
		Data points in (a), (b), and (c) correspond to those in Fig.~\ref{fig:uniform}\,(d), (e) and (f), respectively.
		The results of the nonuniform linearized (solid black), full LLG solution (dashed gray), and micromagnetic simulations (solid red) are shown by lines.
		Shaded areas indicate the uncertainty in the damping time.
	}
\end{figure*}

To quantitatively describe the observed time-, angle- and field-dependent frequency shifts, we adapted the Smit-Suhl approach for linearizing the Landau-Lifshitz-Gilbert (LLG) equation to the case of the laser-driven precession. 
Initially, our augumented macrospin model assumes spatially uniform magnetization dynamics but incorporates the temporal evolution of magnetic parameters following laser excitation. 
Based on previous works~\cite{carpene2010ultrafast,gerevenkov2021effect}, we assume an 
instantaneous temperature increase from its equilibrium value followed by exponential relaxation with time constant $\tau_r$.
The values of $M_S$ were reconstructed from the experimental temperature dependence~\cite{crangle1971magnetization}.
For the evolution of the cubic anisotropy parameter $K_C$, we use a power law~\cite{zener1954classical,akulov1936quantentheorie} relating $K_C(t)$ to $M_S(t)$, where the relative change in $K_C$ scales as the 10th power of the relative change in $M_S$.

The equilibrium saturation magnetization was taken as $M_S (\infty) = 1.72$~MA~m$^{-1}$ for bulk single-crystalline iron~\cite{cullity2011introduction}.
The cubic anisotropy constant $K_C (\infty)$ was determined from the fitting procedure described below.
We fitted the precession frequencies at the selected time delays using the Smit-Suhl formula, varying the relaxation time $\tau_{r}$, the maximum pump-induced temperature increase $\Delta T (0)$, the cubic anisotropy constant $K_C (\infty)$, and the value of uniaxial anisotropy $K_U$.
The best fit, shown by the lines in Fig.~\ref{fig:uniform}\,(b,c), was achieved with $\tau_{r} = 0.95$\,ns, $\Delta T (0) = 392$\,K, $K_C (\infty) = 37.4$\,kJ~m$^{-3}$, and $K_U = 83.6$\,kJ~m$^{-3}$.
Here $\Delta T$ is the temperature increase after initial equilibration of electrons, spins, and lattice temperatures~\cite{CHEN2025}.
The Gilbert damping constant, obtained from the best fit, is $\alpha_G = 4 \cdot 10^{-3}$, which is close to the reported values for epitaxial iron films on MgO~(001) substrate~\cite{khodadadi2020conductivitylike}.
A small increase in the $\alpha_G$ can be attributed to the moderate temperature dependence of the intrinsic damping~\cite{bhagat1974temperature}.
This macrospin model successfully captures both the absolute precession frequencies [Fig.~\ref{fig:uniform}~(b)] and their temporal evolution [Fig.~\ref{fig:uniform}\,(c)].

Beyond the frequency analysis, we also extract the damping time $\tau_d$ of oscillations by fitting the precession envelope with an exponential function, $a \exp(-t/\tau_d)$ [Fig.~\ref{fig:uniform}\,(a)]. 
Fig.~\ref{fig:uniform}\,(e) shows the complete azimuthal dependence of $\tau_d$ at $\mu_0 H_\mathrm{ext}=$0.1~mT.
We measure $\tau_d$ as a function of the external magnetic field strength $H_\mathrm{ext}$ for two in-plane orientations: one close to the EA [$8 \pm 2.5\,^\circ$, Fig.~\ref{fig:uniform}\,(f)] and one close to the HA [$53 \pm 2.5\,^\circ$, Fig.~\ref{fig:uniform}\,(d)]. 
In the experiment, decreasing $H_\mathrm{ext}$ results in an increase in $\tau_d$ in the EA geometry and an anomalous decrease in $\tau_d$ in the HA-geometry.

To comprehend this behavior, we compared our experimental results with the damping times predicted by the Smit-Suhl approach [lines in Fig.~\ref{fig:uniform}\,(d-f)]. 
The comparison reveals a significant discrepancy. 
The theory consistently overestimates the damping time across the entire range of external magnetic field directions for values below 0.2\,T. 
The most pronounced qualitative discrepancy appears when the field is directed near the HA [shaded area in Fig.~\ref{fig:uniform}\,(d)]. 
This geometry corresponds to the model case of a field-induced second-order spin-orientation transition, where the magnetization is forced to align along the hard axis. 
This striking deviation between experiment and the standard macrospin approach motivates a more detailed investigation.

The macrospin approach clearly fails in describe laser-induced precession, as it disregards the finite excitation profile that yields inhomogeneous transient magnetization dynamics~\cite{mishra2023dynamics,gareev2025strongly,davies2019} and also neglects magnetic-dipolar fields~\cite{khokhlov2024spatial}.
To discriminate between these two effects, we first extend the Smit-Suhl approach beyond the single-macrospin approximation to account for the finite sizes of the pump and probe areas. 
The pump pulse creates a Gaussian profile of laser-induced transient heating across the film plane, resulting in a spatial distribution of magnetic parameters. 
The probe pulse, in turn, measures the response with its own Gaussian sensitivity profile. 
Azimuthal symmetry allows us to reduce the problem to a single radial integral from the center of excitation to approximately three standard deviations, beyond which the changes have a negligible effect:


\begin{multline}\label{eq:integral}
    \tilde m_{z} \left[ \Delta T (0),\tau_{r},t \right] = 
    \int_0^{3 \sigma_{pump}} r G(\sigma_{probe}, r) \times \\ m_{z} \left[ \Delta T (0) G(\sigma_{pump}, r) e^{-t/\tau_{r}} \right] dr,
\end{multline}
where $\tilde m_{z}$ is the predicted integral out-of-plane magnetization component reproducing the measured one, $m_{z}$ is the local out-of-plane magnetization obtained from the linearized or full LLG equation, and $G(\sigma, r)$ is a Gaussian function with standard deviations $\sigma_{pump}$ and $\sigma_{probe}$ for the pump and probe spots, respectively.
We neglect the exchange interaction, since for iron films of comparable thickness it does not contribute to the dispersion of spin waves with wavelengths on the order of a micron and larger~\cite{filatov2022spectrum}.
To implement this model, we first calculate a set of precessing magnetizations $m_{z}(t)$ with magnetic parameters corresponding to the local laser-induced heating.  
Integrating this set according to Eq.~(\ref{eq:integral}) yields the predicted signal $\tilde m_{z}(t)$, which inherently includes the interference of oscillations at various radial distances within the excitation spot, weighted by the probe sensitivity. 
By fitting the frequency of the resulting $\tilde m_{z}(t)$ to the experimental data, we refine the estimates of the model parameters. 

To compute the local magnetization $m_{z}$, we use both the linearized and the full LLG equations. 
The best fit using Eq.~(\ref{eq:integral}) yields two sets of parameters.
For the linearized LLG (black solid lines in Fig.~\ref{fig:nonuniform}), we obtain $\tau_{r} = 0.98$\,ns, $\Delta T (0) = 432$\,K, $K_C (\infty) = 37.9$\,kJ~m$^{-3}$, and $K_U = 85.5$\,kJ~m$^{-3}$.
For the full LLG (gray dashed lines in Fig.~\ref{fig:nonuniform}), we obtain $\tau_{r} = 0.74$\,ns, $\Delta T (0) = 344$\,K, $K_C (\infty) = 41.8$\,kJ~m$^{-3}$, and $K_U = 102.2$\,kJ~m$^{-3}$.
Both approaches capture the azimuthal and field dependences of the precession frequency well (not shown); therefore, we focus on the damping time $\tau_d$ in the following.

\begin{figure}
	\includegraphics[width= 0.8\linewidth]{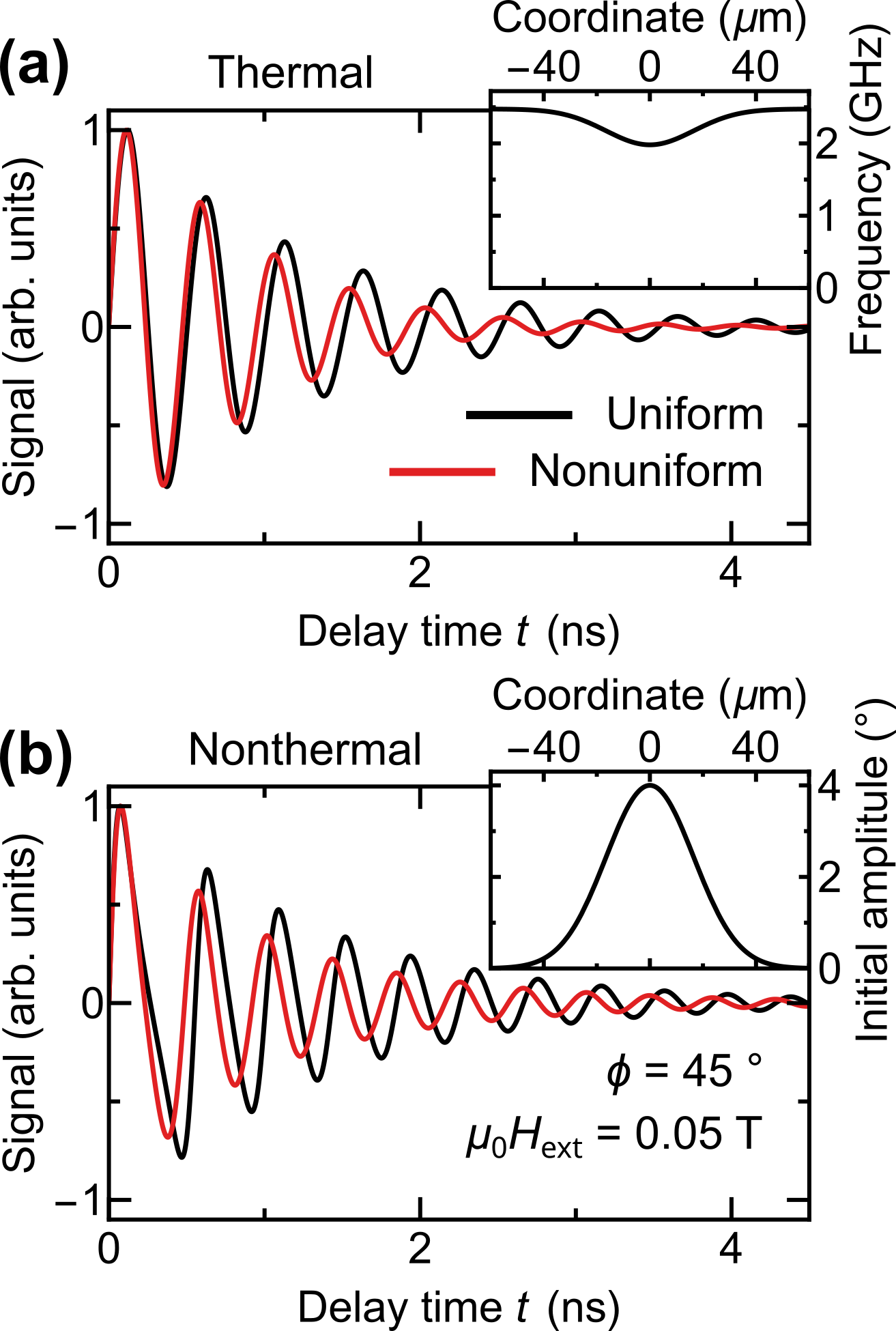}
	\caption{\label{fig:interference}
		{\bf Influence of signal interference from different points across the excitation area on the observed damping time and precession envelope.}
		The simulated signal $m_{z}$ at the center of the pump area and the predicted measured signal $\tilde m_{z}$ obtained using Eq.~(\ref{eq:integral}) are shown for two cases: (a) dispersion of the eigen frequency due to heating (inset shows the frequency dependence on the coordinate); 
		(b) nonlinear shift of the frequency due to inhomogeneity of the initial amplitude  (inset shows the coordinate dependence of the initial amplitude).
	}
\end{figure}

The azimuthal dependence [Fig.~\ref{fig:nonuniform}\,(b)] shows that accounting for the finite sizes of the pump and probe areas leads to a significant reduction in the measured $\tau_d$ values in a narrow range of angles around the HA and in a wide range around the EA.
In sufficiently high fields, the anomalous contributions to decay time are known to vanish~\cite{walowski2008intrinsic,capua2015determination}, which is consistent with our results for fields above 0.2\,T [Fig.~\ref{fig:nonuniform}\,(a)].
However, Figs.~\ref{fig:uniform}\,(b,e)~and~\ref{fig:nonuniform}\,(b) demonstrate that for certain directions of the external field (marked by red arrows), corresponding to points where the frequency depends neither on the heating nor on the precession amplitude, there are no anomalous contributions to $\tau_d$ even at fields at least half as low.
Near the second‑order spin‑reorientation transition [Fig.~\ref{fig:nonuniform}\,(a)], both the linearized and the full LLG predict an anomalous increase in damping time.
However, the full LLG predicts a more pronounced effect and the existence of a local maximum near the transition (marked by black arrows), which better agrees with the experimental data.

This anomalous increase in $\tau_d$ arises from time‑domain interference of the precessing magnetizations within the inhomogeneously excited area.
Specifically, laser‑induced heating across this region creates a spatial distribution of precession frequencies.
Although all precessions start with zero phase difference, the spatial gradient of magnetic parameters causes their phases to diverge over time.
To illustrate this, we show in Fig.~\ref{fig:interference}\,(a) an exponentially decaying sine signal $m_z$ and corresponding $\tilde m_z$ obtained using Eq.~(\ref{eq:integral}) with a spatial frequency variation of 20\,\% (inset).
Note that defects can produce a similar contribution, as reported in~\cite{walowski2008intrinsic,chen2012spin}.

The full LLG introduces an additional nonlinear contribution to the damping, that is not related to the inhomogeneous parameters relaxation.
This contribution becomes significant near the spin‑orientation transition and comes from a dependence of the precession frequency on the initial amplitude~\cite{gerevenkov2026}.
Fig.~\ref{fig:interference}\,(b) shows the solution $m_z$ of the full LLG for a macrospin near the spin‑orientation transition together with the signal $\tilde m_z$ obtained from Eq.~(\ref{eq:integral}) without heating, assuming only the spatial dependence of the initial amplitude shown in the inset.
Hence, near the spin‑orientation transition, both thermal and temperature‑independent nonlinear contributions give raise to the anomalous increase in $\tau_d$.
These contributions lead to a non‑exponential envelope, which is reflected in Fig.~\ref{fig:nonuniform} as the shaded uncertainty region.
We note that the effect of nonlinearity has also been observed for non‑thermal transitions ~\cite{Frej2023,solovev2021second}.

\begin{figure}
	\includegraphics[width= 0.8\linewidth]{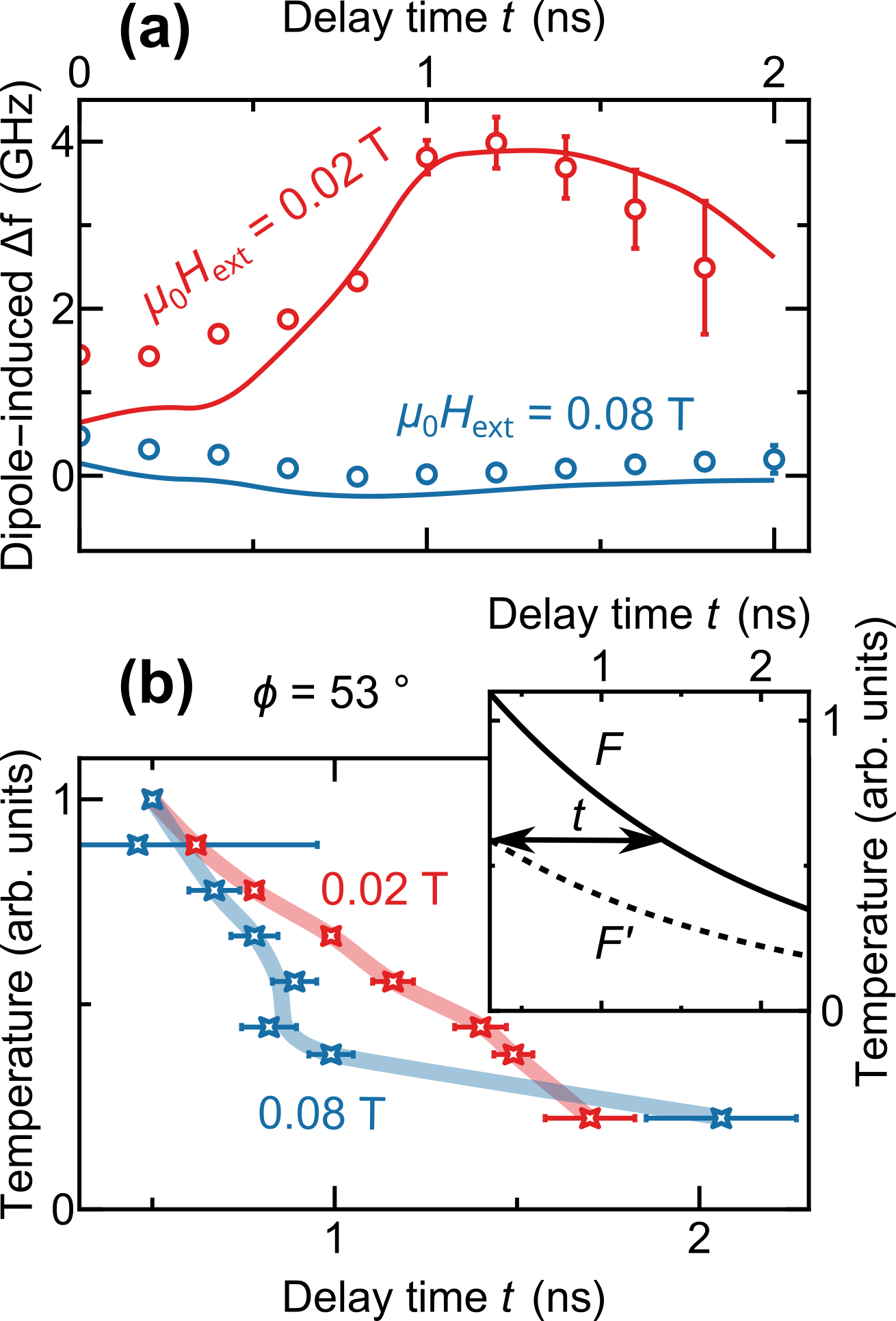}
	\caption{\label{fig:timeDep}
		{\bf Non-monotonous temporal evolution of the laser-induced dipole field contribution.}
		(a) Dipole-induced frequency change $\Delta f$ from the experiment is obtained by subtracting the time‑dependent frequency of the laterally nonuniform LLG model (which accounts for finite pump and probe spot sizes but neglects dipole fields) from the experimental data (points).
		The contribution from micromagnetic simulations is obtained by the same subtraction from the simulation results (lines).
		(b) Effective laser‑induced heating as a function of delay time, derived from the experimental data (see text for details).
		The lines show a guide for the eyes.
		Inset: schematic illustrating the equivalence between a decrease in pump fluence and a shift in measurement time after excitation.
		Data are shown for two external field values: 0.02 (red) and 0.08\,T (blue).
	}
\end{figure}

At this stage, all contributions except for the dipole stray fields arising from local changes in magnetic parameters have been taken into account.
The rigorous consideration of dipole fields presents significant computational challenges. 
It is not only numerically resource-consuming but also breaks the azimuthal symmetry of the problem. 
Moreover, the need to account for their temporal evolution makes an analytical solution virtually impossible. 
To overcome this, we use micromagnetic simulations to compute the estimated signal under experimental conditions and extract frequencies, amplitude and $\tau_d$ as described earlier. 
The best fit was achieved with the parameters $\tau_{r} = 1.75$\,ns, $\Delta T (0) = 748$\,K, $K_C (\infty) = 38.4$\,kJ~m$^{-3}$, and $K_U = 70.6$\,kJ~m$^{-3}$ (red solid lines in Fig.~\ref{fig:nonuniform}).
Compared to the extended macrospin approach, accounting for dipole fields leads to a significant change in the $\Delta T (0)$ and $\tau_{r}$ estimates, which is critical when describing thermal transitions in materials with large $M_S$.
The calculated azimuthal and field dependencies of the damping time $\tau_d$ demonstrate good qualitative and satisfactory quantitative agreement with the experimental results (Fig.~\ref{fig:nonuniform}).
The field dependences show that after $\tau_d$ saturation, a quantitative discrepancy between the predicted and measured values is observed.
This discrepancy was previously observed in epitaxial Fe films~\cite{zhang2020strongly}, where the authors attributed it to anisotropy of $\alpha_G$.
Another possible explanation is the thermal change of $K_U$, which was not considered in this work.

While macrospin Smit-Suhl calculations and numerical LLG simulations often assume an exponential temporal evolution of magnetic parameters, a more accurate description of the experimental data requires accounting for dipole fields, whose temporal evolution is not known {\it a priori}.
To isolate this contribution, we subtract the time-dependent frequency obtained from the numerical LLG solution that accounts for finite pump and probe spot sizes (taken as the most rigorous macrospin approach) from that obtained in micromagnetic simulations and in experiment [lines and symbols in Fig.~\ref{fig:timeDep}\,(a)] for the same parameters.
This procedure reveals that the contribution of dipole fields to the precession frequency is non‑monotonic in time.
Furthermore, this contribution changes its form depending on the magnitude of $H_\mathrm{ext}$.
This finding demonstrates that the temporal evolution of the frequency cannot be adequately described by a simple model of exponential relaxation of magnetic parameters alone, and that a rigorous treatment of dipole fields is essential.


To confirm the impact of non-monotonous dipolar fields experimentally, we developed a method to convert the measured precession frequencies into an effective laser-induced heating scale. 
If the temporal evolution of precession is governed only by the cooling and the corresponding relaxation of $M_S$ and $K_C$, then the following correspondence should hold:  magnetic parameters at a delay $t>0$ after excitation with a pump pulse of a fluence $F$ are equivalent to those at $t=0$ for a pump pulse of fluence $F'<F$ [inset in Fig.~\ref{fig:timeDep}\,(b)]. 
Thus, we used the temporal dependences of the precession frequency obtained in experiments with the largest fluence $F = 38$\,mJ$\cdot$cm$^{-2}$ (from the set of 8 -- 38\,mJ$\cdot$cm$^{-2}$) and mapped the frequencies at various time delays to the frequencies at $t=0$ at lower fluences $F'$. 
The obtained set of $F'$ provides a measure of the decay of the laser-induced heating. 
Fig.~\ref{fig:timeDep}\,(b) shows the resulting temporal evolutions of the heating obtained at different applied magnetic fields in the HA-geometry.
Field-dependent deviation from exponential evolution indicates a breakdown of the initial assumption and reveals the contribution of dipole fields with non-monotonous time dependence, in qualitative agreement with the simulation results [Fig.~\ref{fig:timeDep}\,(a)].

This form changing in the deviation of the effective field from an exponential relaxation explains why the relaxation time $\tau_r$ obtained in our work differs significantly from values reported earlier. 
For instance, in the studies by~\cite{carpene2010ultrafast,gerevenkov2021effect}, complete relaxation of the effective field was not observed within the measurement windows of 0.5~ns and 3~ns, respectively. 
Our results highlight that the overall relaxation process in magnetic systems with pronounced anisotropy is intrinsically linked to the strength and direction of the external field relative to the anisotropy axes. 
Therefore, the contribution of dipole fields must be considered to fully account for the observed anomalous damping.

In conclusion, we have shown that the standard macrospin approach fails to capture the dependence of the precession damping time on the magnitude and direction of the external magnetic field.
Accounting for the finite sizes of the pump and probe spots introduces frequency dispersion across the excitation area, which leads to a measurement‑specific distortion of the precession envelope and a corresponding change in the extracted damping time due to interference of signals from different points within the detection area.
This frequency dispersion originates from two effects: inhomogeneous thermal modification of the magnetic parameters and a non‑thermal nonlinear dependence of the precession frequency on the nonuniform initial amplitude.
Incorporating this dispersion allows us to describe the experimental dependence of the damping time on the magnitude and direction of the external magnetic field, including the anomaly near the spin‑orientation transition, without invoking any change in the intrinsic Gilbert and extrinsic damping.
However, when the magnetization is aligned along certain directions relative to the anisotropy axes, the frequency is nearly insensitive to excitation inhomogeneities, and the measured damping time closely approaches the real value.
Furthermore, in materials with a large saturation magnetization, the dipole fields arising from the finite excitation region contribute to the precession frequency in a non‑monotonic manner that depends on the magnitude and direction of the external field.
Neglecting this contribution leads to a significant underestimation of the laser‑induced heating and the relaxation time.

The work of P.I.G. was supported by the Russian Science Foundation (Grant No.~24-72-00136, https://rscf.ru/project/24-72-00136/).

\bibliography{apssamp}

\end{document}